# Superdeformation: Perspectives and prospects

Jacek Dobaczewski

*Institute of Theoretical Physics, Warsaw University, Hoża 69, PL-00681, Warsaw, Poland*
*Joint Institute for Heavy Ion Research, Oak Ridge National Laboratory, P.O. Box 2008, Oak Ridge, TN 37831, U.S.A.*
*Department of Physics, University of Tennessee, Knoxville, TN 37996, U.S.A.*
*Institut de Recherches Subatomiques, Université Louis Pasteur, Strasbourg I, F-67037 Strasbourg Cedex 2, France*

**Abstract.** We present a review of the mean-field approaches describing superdeformed states, which are currently used and/or being developed. As an example, we discuss in more details the properties of superdeformed $A{\sim}60$ nuclei, and present results of calculations for the rotational band in the doubly magic superdeformed nucleus $^{32}$S.

## INTRODUCTION

Physics of superdeformed nuclear shapes is already a mature field, and a number of excellent review articles exist [1–4]. For the recent developments in experiment we refer the reader to other contributions to this conference. In theory, after the early calculations, which have predicted the existence of the superdeformed shape, see Ref. [2] for the review, and its experimental discovery in 1985 [5], numerous groups have developed sophisticated techniques to perform calculations allowing the description of rotational bands for very elongated shapes. At present, one witnesses a very impressive activity in this field. Almost all these theoretical approaches use the mean-field concept of the nucleus. One can say that the phenomenon of superdeformation constitutes a spectacular manifestation of the mean-field properties of nuclei.

In the present contribution, we review the mean-field methods presently used to describe the rotating superdeformed states, and discuss results obtained in recent years. As an example, we present in more details the calculations for nuclei in the $A{\sim}60$ region, and for $^{32}$S.

## WHAT HAS BEEN DONE IN THEORY AFTER 1995

In this section we attempt at describing the present status of the mean-field theory of superdeformed states. In Table 1 we enumerate groups working in this domain in 1998. They are listed according to types of the mean field used, which are: the Nilsson (NS) [107,108] or Woods-Saxon (WS) [109,110,108] phenomenological mean fields, the non-relativistic self-consistent Skyrme [111–113,108] and Gogny [114,115,108] mean fields, or the self-consistent Relativistic Mean Field (RMF) [116]. As shown in Table 1, calculations are performed either by neglecting the pairing correlations (none), or by including them at various degrees of sophistication, ranging from the simplest $\Delta(\omega)$ method [117,118], or the Bardeen-Cooper-Schrieffer (BCS) or Bogolyubov (HFB) approximations [108], to the Lipkin-Nogami (LN) [119–124] or the particle-number-projection (PNP) [108] methods. Approaches which do include pairing correlations use various interactions in the particle-particle channel. These can be: the seniority ($P^+P$) or quadrupole pairing ($Q^+Q$) forces, the zero-range or zero-range and density-dependent forces, which induce the volume (Vol.) or surface (Surf.) pairing correlations, respectively, or the particle-particle force given by the finite-range Gogny interaction. Aiming at covering only the most recent studies, and assuming the review in Ref. [3] as the baseline, we quote in Table 1 references to papers published since 1995 till the present day.

Ingemar Ragnarsson and collaborators [6–22] have published numerous analyses of superdeformed bands, providing theoretical interpretation at the same time as many initial experimental discoveries have been made. In their calculations they have been minimizing the total energy (obtained within the Strutinsky prescription) with respect to deformation. Depending on the physical situation, pairing correlations have been included within the NS+HFB approach (self-consistent treatment of pairing within a fixed set of NS single-particle energies). Some studies required a given configuration to be followed across the $\epsilon$-$\gamma$ plane, and then the pairing correlations have been neglected, while the level crossings have been removed by the diabatic procedure [125]. In this way, in the $A{\sim}60$ region of superdeformed nuclei, several bands could be interpreted in terms of the band-termination phenomenon [9,15,19,20]. Following the statistical analysis of experimental data [126], which aimed at quantifying the frequency of occurrence of the identical bands, Karlsson, Ragnarsson et al. presented an analogous analysis of the calculated moments of inertia [16,18]. The same group has also performed the analysis of relative quadrupole moments [21] and confirmed their additivity (obtained originally within the HF method [84]), as well as studied the quadrupole polarization charges in terms of the harmonic oscillator (HO) single-particle quantum numbers. Recently, Afanasjev, Ragnarsson, and Ring [22] have presented a joint Nilsson and RMF study of several superdeformed nuclides in the $A\sim60$ region.

**TABLE 1.** 1998 Who Is Who in the mean-field theory of superdeformed states.

| | Refs. | Mean Field | | | | | Pairing | | | | | Pairing Force | | | | |
|---|---|---|---|---|---|---|---|---|---|---|---|---|---|---|---|---|
| | | NS | WS | Skyrme | Gogny | RMF | none | $\Delta(\omega)$ | BCS | HFB | LN | PNP | $P^+P$ | $Q^+Q$ | Vol. | Surf. | Gogny |
| I.Ragnarsson et al. | [6–22] | X | | | | | X | | | X | | | X | | | | |
| T.Nakatsukasa[a] et al. | [23–31] | X | | | | | X | X | | | | | X | | | | |
| Y.Sun[b] et al. | [32–35] | X | | | | | | | X | | | | X | X | | | |
| R.Wyss et al. | [36–57] | | X | | | | | | X | X | X | | X | X | X | | |
| T.R.Werner et al. | [58–65] | | X | | | | X | | | | | | | | | | |
| R.R.Chasman et al. | [66–68] | | X | | | | X | | | | | | | | | | |
| P-H.Heenen[c] et al. | [69–79] | | | X | | | | | X | X | X | | X | | | X | X |
| J.Dobaczewski et al. | [80–89] | | | X | | | X | | | | | | | | | | |
| K.Matsuyanagi et al. | [90] | | | X | | | X | | | | | | | | | | |
| L.Egido et al. | [91–93] | | | | X | | X | | | X | X | X | | | | | | X |
| M.Girod[d] et al. | [94–96] | | | | X | | | | | X | X | | | | | | | X |
| P.Ring et al. | [97–103] | | | | | X | X | | X | | | | X | | | | |
| H.Madokoro et al. | [104–106] | | | | | X | X | | | | | | | | | | |

[a] Plus collective RPA correlations.
[b] Plus the angular momentum projection and coupling to selected multi-qp states. No cranking.
[c] Plus collective GCM correlations.
[d] Plus collective equation of motion.

Takashi Nakatsukasa and collaborators [23–31] have studied superdeformed bands in the $A\sim190$ (and $A\sim150$ [23]) region of nuclei, with the particular emphasis on the collective excitations. They have been solving the RPA equations in the rotating frame and have used separable multipole interactions in doubly-stretched coordinates [127]. Pairing correlations have either been neglected [23], or included within the $\Delta(\omega)$ method. They have proposed an interpretation of several excited superdeformed bands in terms of collective quadrupole or octupole excitations.

Yang Sun and collaborators [32–35] have used the Projected Shell Model (PSM) [128,32] to describe the superdeformed bands in several nuclei. In two aspects this model goes beyond the mean-field approximation. Firstly, it uses the BCS wave functions projected on good angular momentum. (A restriction to axial shapes allows to perform the projection by doing one-dimensional integrals [108].) Secondly, the PSM allows for a mixing of the mean-field ground state with low-lying two- and four-quasiparticle states (in case of an even-even system). The model does not use the cranking approximation, i.e., the high-angular momentum states are obtained by projecting appropriate components from non-rotating states. A diagonalization in such a basis allows to some extent to take into account the configuration changes with increasing spin. Conservation of the angular momentum removes known disadvantages of the cranking approximation which fails when two configurations cross each other. On the other hand, large spaces are probably required to accommodate rotation-induced modifications of states when they are treated in a non-rotating basis (especially in the pairing channel). By conserving the angular momentum, the model also allows to calculate the transition matrix elements in a consistent way, although for large deformations the standard intrinsic-frame treatment is probably precise enough. The model certainly deserves further attention, especially when it will be enriched by the particle-number projection, which can by crucial in a weak-pairing regime typical for the rotating superdeformed states. The model is very successful in describing properties of superdeformed bands in $A\sim190$ nuclei [33].

Ramon Wyss and collaborators [36–57] have provided theoretical interpretation of a large number of experimentally observed superdeformed bands in the $A\sim130$, 150, and 190 regions nuclei. They have used the WS+HFB approach (self-consistent treatment of pairing within a fixed set of WS single-particle energies) together with the LN method to restore the numbers of particles. The total routhians, obtained by the Strutinsky procedure, have been minimized with respect to the $\beta_2$, $\gamma$, and $\beta_4$ deformations. The method constitutes by now a standard approach to describe rotational states, and proved to be extremely successful in reproducing properties of superdeformed bands, see for example Ref. [36].

Tomek Werner and collaborators [58–65] have used a similar method, based on the WS mean field and with no pairing correlations, to describe numerous superdeformed bands in the $A\sim80$ region.

Richard Chasman and collaborators [66–68] have employed another version of the Strutinsky approach, based on the WS mean field, with emphasis on the octupole degree of freedom for nuclei in the $A\sim150$ and 190 regions.

Paul-Henri Heenen and collaborators [69–79] have been using the HFB+Skyrme code in space coordinates, with pairing correlations treated within the LN method and various interactions used in the particle-particle channel. By analyzing superdeformed bands in $A\sim150$ and 190 nuclei they have pointed out the necessity of using the pairing forces having surface character [72]. They have also studied collective correlations at superdeformed shapes by employing the Generator Coordinate Method (GCM) in the quadrupole-octupole [70] and hexadecapole [71] channels. Excitation energies and particle separation energies in secondary minima have also been recently studied [75]. Satuła, Heenen, and collaborators have also performed the joint WS and HF+Skyrme analysis of single-particle alignments in $A\sim190$ nuclei [79].

The present author and collaborators [80–89] have used the HF+Skyrme code constructed in the Cartesian HO basis [85] to describe superdeformed bands in $A\sim60$ and 150 nuclei. The code is fast enough to allow massive calculations for multiple nuclei and configurations. Such results have been obtained for bands around $^{152}$Dy, which allowed to formulate the additivity principle for relative quadrupole moments [84], and to study influence of various time-odd components on properties of superdeformed bands [80,83]. Results for nuclei in the $A\sim60$ region [89] are discussed below.

Masayuki Yamagami and Kenichi Matsuyanagi [90] have recently constructed a new HF+Skyrme code in space coordinates, in which they break all space symmetries. As the first application they present at this conference the calculations for rotational bands in $^{32}$S.

Luis Egido and collaborators [91–93] have studied the superdeformed bands in the $A\sim190$ region by using the HFB method, the Gogny interaction, and the LN and PNP treatments of pairing correlations. They have pointed out the importance of a consistent treatment of the complete interaction in the LN method, and studied this method in the presence of density-dependent terms. In a recent study [93] they have shown that the dynamic moment of inertia of $^{152}$Dy, which in the self-consistent calculations was always slightly larger than in the experiment, cf. Refs. [73,82], comes out just right when the exact PNP calculation is performed.

Michel Girod and collaborators [94–96] have also used the HFB theory and the Gogny interaction, as well as the LN method, and studied the rotational spectra either in the cranking approximation [94,95] or by diagonalizing the collective Bohr Hamiltonian with the inertial functions calculated at zero spin via the Gaussian Overlap Approximation to the GCM [95,96].

Peter Ring and collaborators [97–103] have used the RMF approach without pairing correlations to describe the superdeformed bands in $A\sim140$-150 nuclei [99,102]. They have performed systematic calculations of moments of inertia, quadrupole moments and alignments, and attributed specific configurations to experimentally discovered bands. Similar method has also been used in $A\sim80$ nuclei [97]. By using the BCS pairing correlations (at no spin) they have also analyzed excitation energies of superdeformed minima in $A\sim190$ nuclei [103].

Hideki Madokoro and Masayuki Matsuzaki [104–106] have very recently constructed a cranking code based on the RMF mean field. Their first applications aim at studying the newly discovered superdeformed states in the $A\sim60$ region.

The review presented above is by no means exhaustive, because it covers only the recent theoretical investigations of rotational bands at superdeformation, performed within the mean-field methods. Although these methods constitute at present the mainstream activity, there are also several others which have not been discussed here.

Judging just from the sheer number of studies performed in recent years, one can say that the domain is rich and very active. Within it there are several varieties of approaches, and various groups collaborate and/or compete in achieving the best results.

Close collaboration between the experimental and theoretical groups is a definite asset of the performed studies. In this domain, experiments, interpretations, and predictions are strongly interwoven and mutually supporting, for the benefit of our understanding of the underlying physics.

## SUPERDEFORMATION IN $A\sim60$ NUCLEI

The recent discovery of the next region where the superdeformed bands exist, the region of nuclei with $A\sim60$, see contributions to this conference in Refs. [129,130], has prompted a number of theoretical investigations [9,15,19,20,22,86,89,105,106]. Structure of these nuclei can easily be understood by analyzing the simple HO mean field. In Fig. 1 we present the spherical (1:1) and superdeformed (2:1) HO spectra in light nuclei. Energies of the spherical states read

$$\epsilon_{n_x n_y n_z} = \hbar\omega_0(n_x + \tfrac{1}{2}) + \hbar\omega_0(n_y + \tfrac{1}{2}) + \hbar\omega_0(n_z + \tfrac{1}{2}) = \hbar\omega_0(N_0 + \tfrac{3}{2}), \quad (1)$$

where ($n_x n_y n_z$) are the standard Cartesian HO quantum numbers, and $N_0 = n_x + n_y + n_z$ is the total number of quanta, i.e., the principal HO quantum number. Energies of the superdeformed HO oscillator can be grouped in two families having even and odd numbers of quanta $n_z$ in the $z$ direction, respectively, see Refs. [131,132] and references cited therein, i.e.,

$$\epsilon_{n_x n_y n_z} = \hbar\omega_0'(n_x + \tfrac{1}{2}) + \hbar\omega_0'(n_y + \tfrac{1}{2}) + \tfrac{1}{2}\hbar\omega_0'(n_z + \tfrac{1}{2})$$
$$= \begin{cases} \hbar\omega_0'(N_0' + \tfrac{3}{2}); & \text{for } n_z = 2n_z' \\ \hbar\omega_0'(N_0' + \tfrac{3}{2}) + \tfrac{1}{2}\hbar\omega_0'; & \text{for } n_z = 2n_z' + 1 \end{cases}. \quad (2)$$

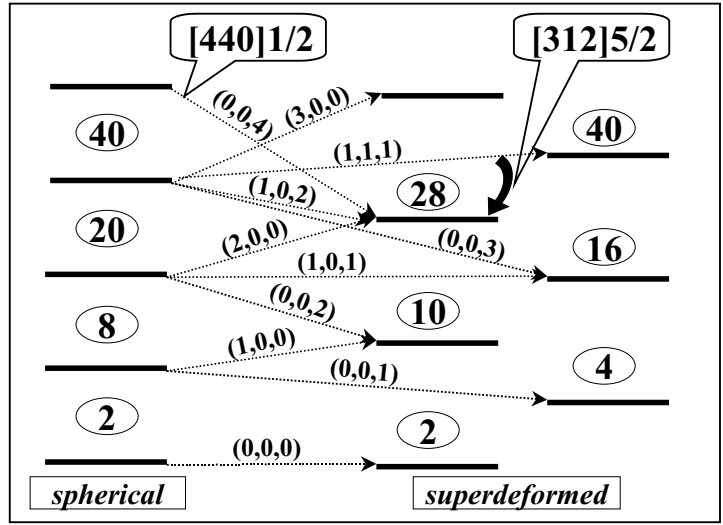

**FIGURE 1.** Spectra of the spherical (axis ratio 1:1, left) and superdeformed (axis ratio 2:1, right) harmonic oscillator (HO). Magic numbers, including two-fold spin degeneracy, are shown in ovals. Dotted arrows show how the states with given Cartesian quantum numbers ($n_x n_y n_z$) move when the HO potential is deformed from the spherical to superdeformed shape. At every arrow only one state is indicated, although several other ones may move in the same way, for example, states (2,0,1) and (0,2,1) move together with (1,1,1). Nilsson quantum numbers of two important orbitals are shown explicitly, namely, [440]1/2 indicates the positive-parity intruder $N_0$=4 orbital, while [312]5/2 shows the only orbital which is pushed down by the spin-orbit interaction below the HO magic gap for 28 particles. From [89].

Here, $N_0' = n_x + n_y + n_z'$ is the principal HO quantum number of each of the two spherical-like HO spectra, and $\omega_0'$ is the corresponding new frequency which is simply related to $\omega$ by the volume-conservation condition.

Dotted arrows in Fig. 1 indicate the correspondence between spherical and superdeformed HO states. For example, the three degenerate $N_0$=1 spherical states, ($n_x n_y n_z$)=(1,0,0), (0,1,0), and (0,0,1), split and move to either one of the two families. Namely, states (1,0,0) and (0,1,0) join the $N_0'$=1 shell of the lower family, while state (0,0,1) forms the $N_0'$=0 shell of the higher family.

One can see that the standard sequence of the spherical magic HO numbers, 2, 8, 20, 40,..., is at the superdeformed shape replaced by the sequence 2, 4, 10, 16, 28, 40.... The numbers in this sequence are either doubled spherical magic numbers or sums of two consecutive spherical magic numbers. Had the HO mean-field been exactly realized in nature, one would therefore observe doubly-magic superdeformed states in $^8$Be, $^{20}$Ne, $^{32}$S, $^{56}$Ni, $^{80}$Zr, and so on.

The simple HO picture can very well be associated with the cluster states in light nuclei [131], for example, with the $\alpha$-$\alpha$ cluster ground state of $^8$Be or the $^{16}$O-$\alpha$ cluster state in $^{20}$Ne. The next candidate in the sequence is, as yet unobserved, the $^{16}$O-$^{16}$O cluster (or superdeformed) state in $^{32}$S (see the next section). Still higher in the sequence one has to take into account the spin-orbit effects which perturb the simple HO mean field. In $A\sim60$ nuclei, the influence of the spin-orbit force is still relatively weak, and amounts to shifting *only one* Nilsson orbital, [312]5/2, below the superdeformed gap. As a result, we observe the superdeformed structures mostly in the nuclei slightly heavier than $^{56}$Ni [129], and we expect the nucleus $^{60}$Zn to be the doubly magic superdeformed core. In Fig. 1, the Nilsson orbital [312]5/2 is marked by a thick arrow.

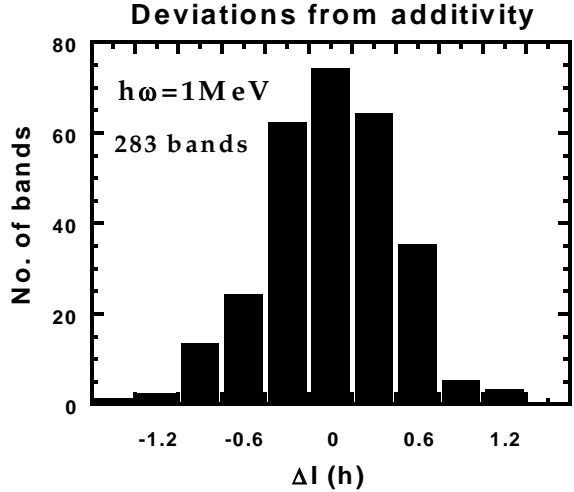

**FIGURE 2.** Histogram of deviations between the HF values of angular momenta calculated at $\hbar\omega=1\,\mathrm{MeV}$ and the values given by the additivity formula (3).

Another very important orbital in this region is the [440]1/2=(0,0,4) intruder state, also explicitly indicated in Fig. 1. In fact, together with the next higher intruder state, [431]3/2, these are the only positive-parity states near the Fermi level of $A\sim60$ superdeformed nuclei. Moreover, since these two orbitals are strongly deformation-driving, one obtains significant changes of deformation between states differing in numbers of occupied $N_0=4$ routhians [86]. Therefore, it is useful to label the superdeformed configurations in this region as $4^n4^p$, where $n$ and $p$ are the numbers of the occupied $N_0=4$ routhians [88]. For example, the doubly magic superdeformed configuration in $^{60}$Zn can be labelled as $4^24^2$. Interestingly, the best theoretical scenario to describe the pair of identical bands in $^{56}$Ni and $^{58}$Cu involves configurations with *different* numbers of occupied $N_0=4$ states [130].

In view of a rapid increase in the amount and quality of experimental data on superdeformed states in $A\sim60$ nuclei, one can aim at systematic studies of their properties. Theoretically, such systematic investigations indicate [89] that the quadrupole moments of nuclei in this region, similarly as those in $A\sim150$ nuclei [84], very well obey [86] the additivity principle. In the present paper we analyze the similar hypothesis for the relative angular momenta, namely, we suppose that at a given angular frequency $\omega$, the angular momentum of a given $A\sim60$ superdeformed configuration can be described through the additive contributions,

$$I_{\mathrm{add}}(\omega) = I\left[^{60}\mathrm{Zn}\right](\omega) + \sum_p \delta I_p(\omega)n_p + \sum_h \delta I_h(\omega)n_h, \qquad (3)$$

with respect to the $^{60}$Zn core. In Eq. (3), $I\left[^{60}\mathrm{Zn}\right]$ denotes the angular momentum of the doubly magic $4^24^2$ configuration in $^{60}$Zn, while $n_p$ and $n_h$ denote respectively the occupation numbers (equal 0 or 1) of particles and holes which have to be created in the $^{60}$Zn core in order to obtain the given configuration in an adjacent nucleus.

Individual contributions of particles ($\delta I_p$) and holes ($\delta I_h$) can be obtained by fitting formula (3) (separately at each value of $\omega$) to the calculated HF angular momenta of a large set of configurations. Details of this procedure will be given elsewhere [89]. As an example, in Fig. 2 we give the histogram of deviations,

$$\Delta I(\omega) = I_{\mathrm{HF}}(\omega) - I_{\mathrm{add}}(\omega), \qquad (4)$$

between the HF results and the additivity formula (3), obtained at $\hbar\omega=1\,\mathrm{MeV}$. One can see that the additivity hypothesis for the angular momenta does not work so precisely as it does for the quadrupole moments. Typical deviations turn out to be of about $0.5\,\hbar$, which is large compared to the precision of relative alignments we want to describe. Therefore, in all particular cases, measured relative alignments should be compared with calculated values for a given particular pair of configurations.

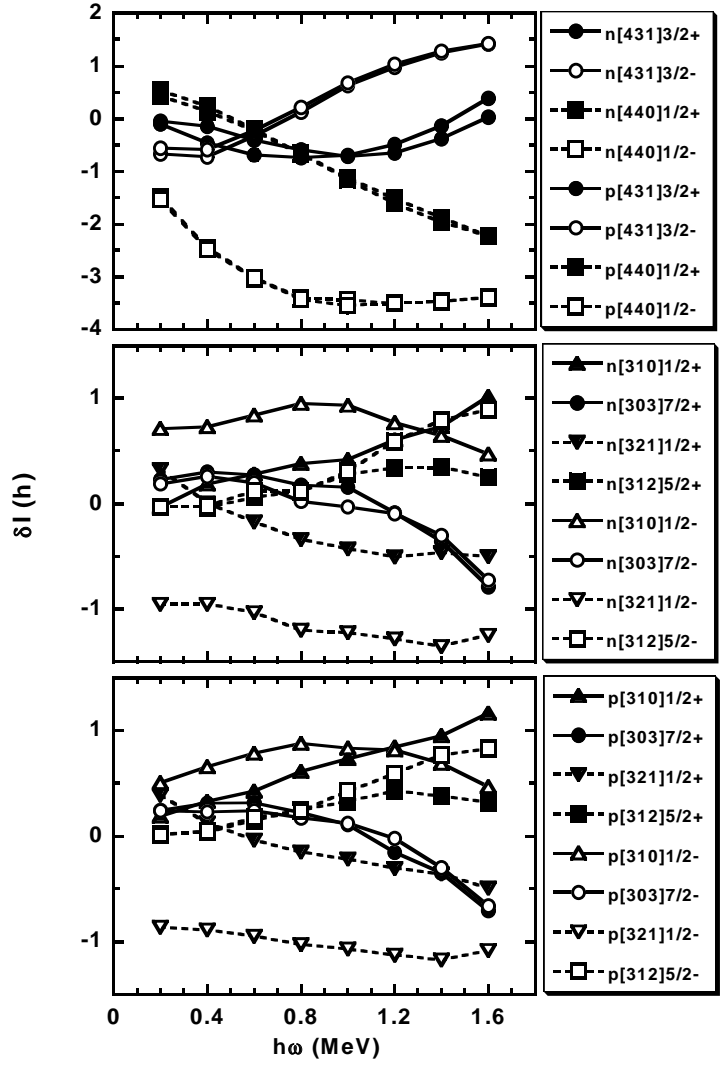

**FIGURE 3.** Contributions of particle and hole orbitals (shown with solid and dashed lines, respectively) to angular momenta in $A{\sim}60$ superdeformed nuclei. Closed and open symbols correspond to the signature quantum numbers $r=+i$ and $r=-i$, respectively.

However, the obtained contributions of individual routhians, shown in Fig. 3, may serve as indicators of which orbital is most important in providing the angular momentum to the nucleus. As seen in the top panel of Fig. 3, the intruder $N_0=4$ orbitals carry much larger angular momenta than the negative-parity states shown in the lower two panels (note the difference of scale). Contributions of proton and neutron intruder orbitals are almost equal. One should also note that beyond $\hbar\omega\simeq 0.6\,\mathrm{MeV}$, the proton and neutron Nilsson routhians [440]1/2− carry almost constant alignments. Therefore, they may lead to identical bands in nuclei which *differ* by the occupation of this particular orbital. Similarly, the negative-parity orbitals [321]1/2−, which carry constant alignments of about $-1\hbar$ may also lead to identical bands.

Results shown in Fig. 3 may also be useful in giving the changes of dynamic moments of inertia $\mathcal{J}^{(2)}(\omega)=dI(\omega)/d\omega$, which arise when a given orbital becomes occupied. Indeed, slopes of curves presented in Fig. 3 are equal to relative changes of the dynamic moments of inertia. From these results it is clear that the values of $\mathcal{J}^{(2)}$ should not depend on the occupation of the first intruder routhian [440]1/2−, and should increase with occupying the following three intruder routhians [440]1/2+, [431]3/2−, and [431]3/2+ (note that the *hole* contribution of [440]1/2+ is shown in Fig. 3).

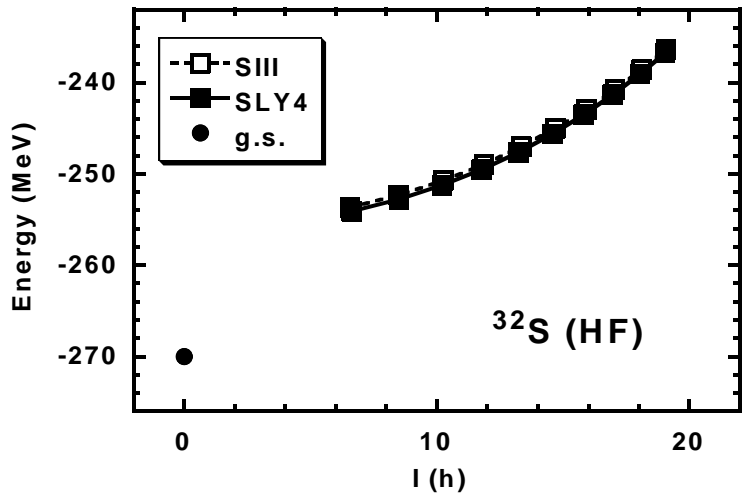

**FIGURE 4.** Energies of the superdeformed band in $^{32}$S as functions of spin, calculated within the HF approximation for the SIII (open squares) and SLy4 (full squares) Skyrme interactions. The dot indicates the calculated HF energy of the ground-state configuration.

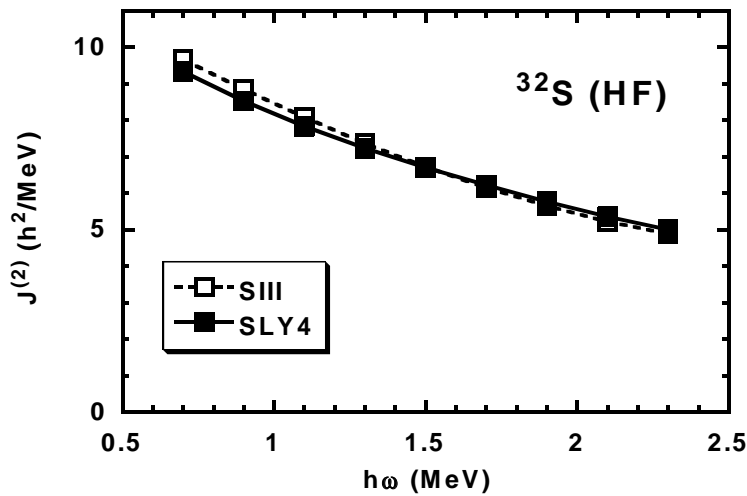

**FIGURE 5.** Same as in Fig. 4 for the dynamic moments of inertia as functions of the rotational frequency.

## THE LAST CHALLENGE: $^{32}$S

As illustrated in Fig. 1, the nucleus $^{32}$S (16 protons and 16 neutrons) is the missing link between the known region of superdeformed nuclei around $^{60}$Zn and the cluster-like structures in lighter nuclei, like in $^{20}$Ne and $^{8}$Be. In fact, the first indication that the cluster states in $^{32}$S may exist is provided by the measurements in the $^{16}$O-$^{16}$O breakup channel [133]. On the other hand, numerous mean-field calculations, both non-self-consistent [134] and self-consistent [135], as well as the $\alpha$-cluster calculations [136], predict in this nucleus the existence of the 2:1 deformed structures. Such states should coexist with numerous low-deformation states already known in this nucleus [137], which are very well described by the $sd$-shell model [138].

In Figs. 4 and 5 we show results of the HF calculations performed with the Skyrme SIII [139] and SLy4 [140,141] interactions. Both interactions give very similar predictions for the superdeformed band which is based on the $3^2 3^2$ configuration, i.e., with altogether four $N_0$=3 intruder states (originating from the $1f_{7/2}$ spherical orbital) occupied. The band should begin around $I$=6$\hbar$ and continue beyond the spin of 20$\hbar$, where it may be crossed by a rapidly descending configuration based on the [440]1/2 orbital, cf. Ref. [90]. The existence of the superdeformed band in $^{32}$S is predicted by these same models and approaches as those successfully used in describing the superdeformed structures in heavier nuclei. Its discovery would therefore constitute a spectacular confirmation of the predictive power of mean-field methods.

## ACKNOWLEDGMENTS

Collaboration, discussions, help, and friendship of my colleagues Witek Nazarewicz and Wojtek Satuła are gratefully acknowledged. This research was supported in part by the Polish Committee for Scientific Research (KBN) under Contract No. 2 P03B 040 14, and by the computational grants from the Interdisciplinary Centre for Mathematical and Computational Modeling (ICM) of the Warsaw University, from the *Institut du Développement et de Ressources en Informatique Scientifique* (IDRIS) of CNRS, France, and from the *Regionales Hochschulrechenzentrum Kaiserslautern*, Germany.